\documentclass{article}
\usepackage{amsfonts,amssymb,amsthm,amsmath}
\usepackage{graphicx}
\usepackage{footnote}
\usepackage{float}
\makesavenoteenv{tabular}
\usepackage{comment}
\usepackage[bookmarks]{hyperref}
\usepackage{authblk}

\theoremstyle{plain}

\theoremstyle{definition}
\newtheorem{deph}{Definition}

\DeclareMathOperator{\e}{\mathcal{E}}

\DeclareMathOperator{\xor}{\veebar}

\DeclareMathOperator*{\argmin}{arg\,min}

\makeatletter
\renewcommand{\maketag@@@}[1]{\hbox{\m@th\normalsize\normalfont#1}}%
\makeatother

\title{\sc Hierarchical Quantification of Synergy in Channels}
\author[1]{Paolo Perrone\thanks{Correspondence: perrone@mis.mpg.de}}
\author[1,2,3]{Nihat Ay}
\affil[1]{\small Max Planck Institute for Mathematics in the Sciences\\ Inselstrasse 22\\ 04103 Leipzig, Germany}
\affil[2]{\small Faculty of Mathematics and Computer Science, University of Leipzig\\ PF 100920\\ 04109 Leipzig, Germany}
\affil[3]{\small Santa Fe Institute, 1399 Hyde Park Road, Santa Fe, NM 87501, USA}
\date{}
\begin{document}
\maketitle
\thispagestyle{empty}
\addcontentsline{toc}{subsection}{Abstract}

\begin{abstract}
The decomposition of channel information into synergies of different order is an open, active problem
in the theory of complex systems. 
Most approaches to the problem are based on information theory, and propose decompositions of
mutual information between inputs and outputs in se\-veral ways, none of which is generally accepted yet.

We propose a new point of view on the topic.
We model a multi-input channel as a Markov kernel. 
We can project the channel onto a series of exponential families which form a hierarchical structure. 
This is carried out with tools from information geometry, in a way analogous to the projections of 
probability distributions introduced by Amari.
A Pythagorean relation leads naturally to a decomposition of the mutual information between 
inputs and outputs into terms which represent 
single node information; pairwise interactions; and in general $n$-node 
interactions.

The synergy measures introduced in this paper can be easily evaluated by an iterative scaling algorithm, which is a standard procedure in information
geometry.

\end{abstract}

\section{Introduction}

In complex systems like biological networks, for example neural networks, a basic principle is that their functioning is based on the correlation and interaction of their different parts.
While correlation between two sources is well understood, and can be quantified by Shannon's mutual
information (see for example \cite{kakihara}), there is still no generally accepted theory for interactions of three nodes or more.
If we label one of the nodes as ``output'', the problem is equivalent to determine how much two (or more) input nodes interact 
to yield the output.
This concept is known in common language as ``synergy'', which means ``working together'', or performing a task that would not be feasible by the single parts separately.

There are a number of important works which address the topic, but the problem is still considered open.
The first generalization of mutual information was  \emph{interaction information} (introduced in \cite{mcgill}), defined for three nodes in terms of the joint and marginal entropies:
\begin{align}\label{ii}
 I(X:Y:Z)= &-H(X,Y,Z) + H(X,Y) + H(X,Z) + H(Y,Z)\, +\notag \\ &- H(X) - H(Y) - H(Z)\;.
\end{align}
Interaction information is defined symmetrically on the joint distribution, but most approaches interpret it by looking at a channel, rather than
a joint distribution, $(X,Y)\to Z$.
For example, we can rewrite \eqref{ii} equivalently in terms of mutual information (choosing $Z$ as ``output''):
\begin{equation}
 I(X:Y:Z)= I(X,Y:Z)-I(X:Z)-I(Y:Z)\;,
\end{equation}
where we see that it can mean intuitively ``how much the whole $(X,Y)$ gives more (or less) information about $Z$ than the sum of the parts separately''.
Another expression, again equivalent, is:
\begin{equation}
 I(X:Y:Z)= I(X:Y|Z)-I(X:Y)\;,
\end{equation}
which we can interpret as ``how much conditioning over $Z$ changes the correlation between $X$ and $Y$'' (see \cite{bialek}).
Unlike mutual information, interaction information carries a sign:
\begin{itemize}
 \item $I>0$: \emph{synergy}. Conditioning on one node \emph{increases} the correlation between the remaining nodes. Or, the whole gives more
 information than the sum of the parts. Example: XOR function.
 \item $I<0$: \emph{redundancy}. Conditioning on one node \emph{decreases}, or \emph{explains away} the correlation between the remaining nodes. Or, the whole gives less 
 information than the sum of the parts. Example: $X=Y=Z$.
 \item $I=0$: \emph{3-independence}. Conditioning on one node has no effect on the correlation between the remaining nodes.
 Or, the whole gives the same amount of information as the parts separately. The nodes can nevertheless still be conditionally dependent. 
 Example: independent nodes.\footnote{For an example in which $I=0$ but the nodes are not independent, see \cite{beer}.} 
\end{itemize}

As argued in \cite{schneidmann}, \cite{beer}, and \cite{griffith}, however, this is not the whole picture. 
There are systems which exhibit both synergetic and redundant behavior, and interaction information only quantifies the \emph{difference}
of synergy and redundancy, with a priori no way to tell the two apart. 
In a system with highly correlated inputs, for example, the synergy would remain unseen, as it would be cancelled by the redundancy.
Moreover, this picture breaks down for more than three nodes. 
Another problem, pointed out in \cite{schneidmann} and \cite{hierarchy}, is that redundancy (as for example in $X=Y=Z$) can be described in terms of pairwise interactions, not triple, 
while synergy (as in the XOR function) is purely threewise. Therefore, $I$ compares and mixes information quantities of different nature.

A detailed explanation of the problem for two inputs is presented in \cite{beer} and it yields a decomposition (``Partial
Information Decomposition, PID) as follows: there exist two non-negative quantities, \emph{Synergy} and \emph{Redundancy}, such that
\begin{equation}
 I(X,Y:Z) = I(X:Z)+I(Y:Z)+Syn - Red\;,
\end{equation}
or equivalently:
\begin{equation}
 I(X:Y:Z) = Syn - Red\;.
\end{equation}
Moreover, they define \emph{unique information} for the inputs $X$ and $X_2$ as:
\begin{align}
 UI(X) &= I(X:Z)- Red\;, \\
  UI(Y) &= I(Y:Z)- Red\;,
\end{align}
so that the total mutual information is decomposed positively:
\begin{equation}\label{pid}
 I(X,Y:Z) = UI(X)+UI(Y) + Red + Syn\;.
\end{equation}
What these quantities intuitively mean is:
\begin{itemize}
 \item Redundancy -- information available in both inputs;
 \item Unique information -- information available only in one of the inputs;
 \item Synergy -- information available only when both inputs are present, ari\-sing purely from their interaction.
\end{itemize}
In this formulation, if one finds a measure of synergy, one can automatically define compa\-tible measures of redundancy
and unique information (and viceversa),
provided that the measure of synergy is always larger or equal to $I(X:Y:Z)$, and that the resulting measure of redundancy
is less or equal than $I(X:Z)$ and $I(Y:Z)$.
Synergy, redundancy, and unique information are defined on a channel, and choosing a different channel with the same joint distribution 
(e.g. $(Y,Z)\to X$) may yield a different decomposition.  

In \cite{griffith} is presented an overview of (previous) measures of synergy, and their shortcomings in standard examples.
In the same paper is then presented a newer measure for synergy, defined equivalently in \cite{entropy} as:
\begin{equation}
 CI(X,Y;Z):=I(X,Y:Z)- \min_{p^*\in \wedge} I_{p^*}(X,Y:Z)\;,
\end{equation}
where $\wedge$ is the space of distributions with prescribed marginals:
\begin{equation}
 \wedge = \big\{ q\in P(X,Y,Z)\;\big|\; q(X,Z)=p(X,Z), q(Y,Z)=p(Y,Z) \big\}\;.
\end{equation}
This measure satisfies interesting properties (proven in \cite{griffith} and \cite{entropy}), which make it compatible with Williams and Beer's PID,
and with the intuition in most examples.
However, it was proven in \cite{rauh} that such an approach can \emph{not} work in the desired way for more than three nodes (two inputs).

Our approach uses information geometry \cite{amari}, extending previous work on hierarchical decompositions \cite{hierarchy} and complexity \cite{complexity}.
(Compare the related approach on information decomposition pursued in \cite{salge}.)
The main tools of the present paper are KL-projections, and the Pythagorean relation that they sa\-tisfy. This allows (as in \cite{hierarchy}) to form
hierarchies of interactions of different orders in a geometrical way. In the present problem, we decompose mutual information between inputs and outputs of 
a channel $k$, for two inputs, as:
\begin{equation}
 I(X,Y:Z)=d_1(k) + d_2(k)\;,
\end{equation}
where $d_2$ quantifies synergy (as in equation \eqref{pid}), and $d_1$ integrates all the lower order terms ($UI,Red$), quantifying the so-called \emph{union information} (see 
\cite{griffith}). One may want to use this measure of synergy to form a complete decomposition analogous to \eqref{pid}, but this does not work, as
in general it is not true that $d_2\le I(X:Y:Z)$. For this reason, we keep the decomposition more coarse, and we do not divide union information into unique and redundant.

For more inputs $X_1,\dots,X_N$, the decomposition generalizes to:
\begin{equation}
 I(X_1,\dots,X_N:Z)=d_1(k)+\dots+d_N(k)=\sum_{i=1}^N d_i(k)\;,
\end{equation}
where higher orders of synergy appear.

Until now, there seems to be no way of rewriting the decomposition of \cite{griffith} and \cite{entropy} in a way consistent with information
geometry, and more in general, Williams and Beer's PID seems hard to write as a geometric decomposition.
A comparison between $d_2$ and the measure $CI$ of \cite{griffith} and \cite{entropy} is presented in Section \ref{jfr}. There we show that $d_2\le CI$,
and we argue, with a numerical example, that $CI$ overestimates synergy at least in one case.

For a small number of inputs $(\lesssim 5)$, our quantities are easily computable with the standard algorithms of information geometry (like iterative scaling \cite{csiszar}).
This allowed to get precise quantities for all the examples considered.

\subsection{Technical Definitions}

We consider a set of $N$ input nodes $V=\{1,\dots,N\}$, taking values in the sets $X_1,\dots,X_N$, and an output node, 
taking values in the set $Y$. We write the input globally as $X:=X_1\times\dots\times X_N$.
For example, in biology $Y$ can be the phenotype, and $X$ can be a collection of genes determining $Y$.
We denote by $F(Y)$ the set of real functions on $Y$, and with $P(X)$ the set of probability measures on $X$. 

We can model the channel from $X$ to $Y$ as a Markov kernel (called also stochastic kernel, transition kernel, or
stochastic map) $k$, that assigns to each $x\in X$ a probability measure on 
$Y$ (for a detailed treatment, see \cite{kakihara}).
Here we will consider only finite systems, so we can think of a channel simply
as a transition matrix (or stochastic matrix), whose rows sum to one.
\begin{equation}\label{stoc}
 k(x;y)\ge 0\quad \forall x,y; \qquad \sum_{y} k(x;y) =1 \quad \forall x\;.
\end{equation}
The space of channels from $X$ to $Y$ will be denoted by $K(X;Y)$.
We will denote by $X$ and $Y$ also the corresponding random variables, whenever this does not lead to confusion.

Conditional probabilities define channels: if $p(X,Y)\in P(X,Y)$ and the marginal $p(X)$ is strictly positive, then $p(Y|X)\in K(X;Y)$ is a well-defined
channel. Viceversa, if $k\in K(X;Y)$, given $p\in P(X)$ we can form a well-defined joint probability:
\begin{equation}
 pk(x,y):= p(x)\,k(x;y)\quad \forall x,y\;.
\end{equation}

An ``input distribution'' $p\in P(X)$ is crucial also
to extend the notion of divergence from probability distributions to channels. 
The most natural way of doing it is the following. 

\begin{deph}
 Let $p\in P(X)$, let $k,m\in K(X;Y)$. Then:
\begin{equation}\label{mkdiv}
 D_p(k||m):= \sum_{x,y} p(x)\,k(x;y)\,\log\dfrac{k(x;y)}{m(x;y)}\;.
\end{equation}
\end{deph}

Defined this way, $D_p$ is affine in $p$.
It is worth noticing that $D_p$ is in general \emph{not} equal to $D(k_*p||m_*p)$. 
Moreover, it has an important compatibility property.
 Let $p,q$ be joint probability distributions on $X\times Y$, and let $D$ be the KL-divergence. Then:
 \begin{equation}
  D(p(X,Y)||q(X,Y)) = D(p(X)||q(X)) + D_{p(X)}(p(Y|X)||q(Y|X))\;.
 \end{equation}

We will now illustrate our geometric ideas in channels with one, two, and three input nodes, then we present some examples.
The general case will be addressed in Section \ref{decomp}.

\section{Geometric Idea of Synergy}\label{idea}

\paragraph{Mutual information as motivation.}
It is a well-known fact in information theory that Shannon's mutual information can be written as a KL-divergence:
\begin{equation}
 I_p(X:Y) = D\big(p(X,Y)||p(X)p(Y)\big)\;.
\end{equation}
From the point of view of information geometry, this can be interpreted as a ``distance'' between the real
distribution and a product distribution that has exactly the same marginals, but maximal entropy.
In other words, we have:
\begin{equation}
 I_p(X:Y) = \inf_{\substack{q\in P(X)\\r\in P(Y)}} D\big(p(X,Y)||q(X) r(Y)\big)\;.
\end{equation}
The distribution given by $p(X)p(Y)$ is optimal in the sense that:
\begin{equation}
 p(X)p(Y) = \argmin_{\substack{q\in P(X)\\r\in P(Y)}} D\big(p(X,Y)||q(X) r(Y)\big) \;.
\end{equation}

\begin{figure}[H]
 \centering
 \includegraphics[scale=0.7,keepaspectratio=true]{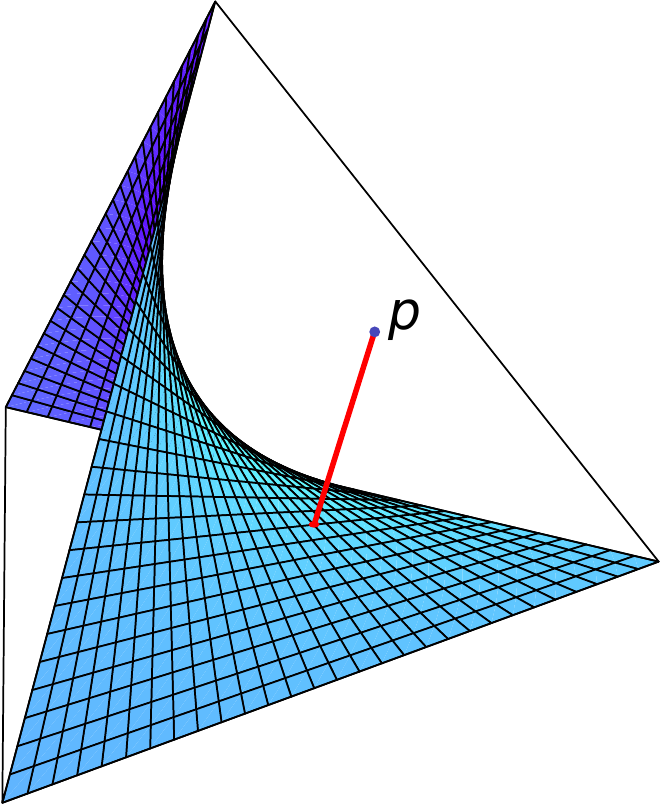}
 \caption{For two binary nodes, the family of product distributions is a surface in a 3-dimensional simplex.}
\end{figure}

The divergence between $p$ and a submanifold is, as usual in geometry, the ``distance'' between $p$ and the ``closest point'' on that submanifold,
which in our case is the geodesic projection w.r.t. the mixture connection.  

\paragraph{Extension to channels.}
We can use the same insight with channels. Instead of a joint distribution on $N$ nodes, we consider a channel from an input $X$ to an output $Y$.
Suppose we have a family $\e$ of channels, and a channel $k$ that may not be in $\e$. Then, just as in geometry, we
can define the ``distance'' between $k$ and $\e$.
\begin{deph}
 Let $p$ be an input distribution. The divergence between a channel $k$ and a family of channels $\e$ is given by:
 \begin{equation}
  D_p(k||\e):=\inf_{m\in\e} D_p(k||m)\;.
 \end{equation}
 If the minimum is uniquely realized, we call the channel
 \begin{equation}
  \pi_{\e}k:=\argmin_{m\in\e} D_p(k||m)\;
 \end{equation}
 the \emph{KL-projection} of $k$ on $\e$ (and simply ``a'' KL-projection if it is not unique). 
\end{deph}
We will always work with compact families, so the minima will
always be realized, and for strictly positive $p$ they will be unique (see Section \ref{decomp} for the details).

We will consider families $\e$ for which the KL-divergence satisfies a Pythagorean equality (see Figure 2 below for some intuition):
\begin{equation}\label{pyt}
 D_p(k||m)=D_p(k||\pi_{\e}k) + D_p( \pi_{\e}k||m)
\end{equation}
for every $m\in\e$.  These families (technically, closures of exponential families) are defined in Section \ref{decomp}.

\begin{figure}[H]
 \centering
 \includegraphics[scale=1,keepaspectratio=true,clip=true,trim=40pt 40pt 0pt 50pt]{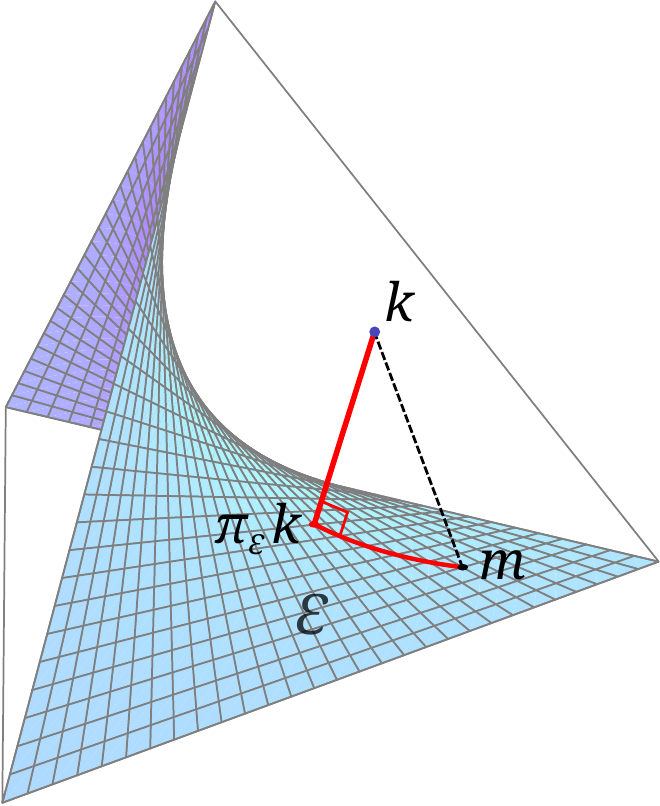}
 \caption{Illustration of the Pythagoras theorem for projections}
\end{figure}

\paragraph{One input.}
Consider first one input node $X$, with input distribution $p(X)$, and one output node $Y$.
A \emph{constant channel} $k$ in $K(X;Y)$ is a channel whose entries do not depend on $X$ (more precisely: $k(x;y)=k(x';y)$ for any $x,x',y$). 
This denomination is motivated by the following properties:
\begin{itemize}
 \item They correspond to channels that do not use any information from the input to generate the output.
 \item The output distribution given by $k$ is a probability distribution on $Y$ which does not depend on $X$.
 \item Deterministic constant channels are precisely constant functions.
\end{itemize}
We call $\e_0$ the family of constant channels. Take now any channel $k\in K(X;Y)$. 
If we want to quantify the dependence in $k$ of $Y$ on $X$ we can then look at the divergence of $k$ from the constant channels:
\begin{equation}
 d_1(k):=  D_p(k||\e_0)\;.
\end{equation}
The minimum is realized in $\pi_{\e_0}k$. We have that:
\begin{align}
 d_1(k)&=D_p(k||\pi_{\e_0}k) = \sum_{x,y}p(x)\,k(x;y)\log\dfrac{k(x;y)}{\pi_{\e_0}k(y)} \\
 &= H_{p\pi_{\e_0}k}(Y) - H_{pk}(Y|X)= I_{pk}(X:Y)\;,
\end{align}
so that consistently with our intuition, the dependence of $Y$ on $X$ is just the mutual information. 
From the channel point of view, it is simply the divergence from the constant channels. 
(A rigorous calculation is done in Section \ref{decomp}.)

\paragraph{Two inputs.}
Consider now two input nodes with input probability $p$ and one output node. We can again define the family
$\e_0$ of constant channels, and the same calculations give:
\begin{equation}\label{mi2}
 D_p(k||\e_0)= I_{pk}(X_1,X_2:Y)\;.
\end{equation}
This time, though, we can say a lot more: the quantity above can be decomposed.
In analogy with the independence definition for probability distributions, 
we would like to define a split channel as a product channel of its parts: $p(y|x_1,x_2)=p(y|x_1)\,p(y|x_2)$.
Unfortunately, the term on the right would be in general not normalized, so we replace the condition by a weaker one.
We call the channel $k(X_1,X_2;Y)$ \emph{split} if it can be written as:
\begin{equation}
 k(x_1,x_2;y) = \phi_0(x_1,x_2)\,\phi_1(x_1;y)\,\phi_2(x_2;y)
\end{equation}
for some \emph{functions} $\phi_0,\phi_1,\phi_2$, which in general are not themselves channels (in parti\-cular, $\phi_i(x_i;y)\ne p(y|x_i)$).
We call $\e_1$ the family of split channels. This family corresponds to those channels that do not have any synergy. This is a special case of
an exponential family, analogous to the family of product distributions of Figure 1. 
The examples ``single node'' and ``split channel'' in the next section belong exactly to this family.
Take now any channel $k(X_1,X_2;Y)$.  In analogy with mutual information, we call \emph{synergy} the divergence:
\begin{equation}\label{syn}
 d_2(k):= D_p(k||\e_1)\;.
\end{equation} 
Simply speaking, our synergy is quantified as the deviation of the channel from the set $\e_1$ of channels without synergy.

We can now project $k$ first to $\e_1$, and then to $\e_0$.
Since $\e_0$ is a subfamily of $\e_1$, the following Pythagoras relation holds from \eqref{pyt}:
\begin{equation}
 D_p(k||\pi_{\e_0}k)=D_p(k||\pi_{\e_1}k)+D_p(\pi_{\e_1}k||\pi_{\e_0}k)\;.
\end{equation}
If in analogy with the one-input case we call the last quantity $d_1$, we get from \eqref{mi2} and \eqref{syn}:
\begin{equation}
 I_{pk}(X_1,X_2:Y) = d_2(k) + d_1(k)\;.
\end{equation}
The term $d_1$ measures how much information comes from single nodes (but it does not tell which nodes).
The term $d_2$ measures how much information comes from the synergy of $X_1$ and $X_2$ in the channel.
The example ``XOR'' in the next section will show this.

If we call $\e_2$ the whole $K(X;Y)$, we get $\e_0\subset \e_1\subset \e_2$ and:
\begin{equation}
 d_i(k):= D_p(\pi_{\e_i}k||\pi_{\e_{i-1}}k)\;.
\end{equation}

\paragraph{Three inputs.}
Consider now three nodes $X_1,X_2,X_3$ with input probability $p$, and a channel $k$. We have again:
\begin{equation}
 D_p(k||\e_0)= I_{pk}(X_1,X_2,X_3:Y)\;.
\end{equation}
This time we can decompose the mutual information in different ways. We can for example look at split channels,
i.e. in the form:
\begin{equation}\label{eone}
k(x_1,x_2,x_3;y)=\phi_0(x)\,\phi_1(x_1;y)\,\phi_2(x_2;y)\,\phi_3(x_3;y) 
\end{equation}
for some $\phi_0,\phi_1,\phi_2,\phi_3$. As in the previous case, we call this family $\e_1$. 
Or we can look at more interesting channels, the ones in the form:
\begin{equation}
k(x_1,x_2,x_3;y)=\phi_0(x)\,\phi_{12}(x_1,x_2;y)\,\phi_{13}(x_1,x_3;y)\,\phi_{23}(x_2,x_3;y) 
\end{equation}
for some $\phi_0,\phi_{12},\phi_{13},\phi_{23}$. We call this family $\e_2$, and it is easy to see that:
\begin{equation}
 \e_0\subset \e_1\subset \e_2\subset\e_3\;,
\end{equation}
where $\e_0$ denotes again the constant channels, and $\e_3$ denotes the whole $K(X;Y)$.
We define again:
\begin{equation}
 d_i(k):= D_p(\pi_{\e_i}k||\pi_{\e_{i-1}}k)\;.
\end{equation}
This time, the Pythagorean relation can be nested, and it gives us:
\begin{equation}\label{series}
 I_{pk}(X_1,X_2,X_3:Y) = d_3(k) + d_2(k) + d_1(k)\;,
\end{equation}
The difference between pairwise synergy and threewise synergy is shown in the ``XOR'' example in the next section.

Now that we have introduced the measure for a small number of input, we can study the examples from the literature \cite{griffith}, and
show that our measure is consistent with the intuition.
The general case will be more in rigor introduced in Section \ref{decomp}.

\section{Examples}\label{examples}

Here we present some examples of decomposition for well-known channels. 
All the quantities have been computed using an algorithm analogous to iterative scaling (as in \cite{csiszar}).

\paragraph{Single Node Channel.}

The easiest example is considering a channel which only depends on $X_1$, i.e.:
\begin{equation}
 I(X:Y) = I(X_1:Y)\;. 
\end{equation}

For example, consider 3 binary input nodes $X_1,X_2,X_3$ with constant input probability, 
and one binary output node $Y$ which is an exact copy of $X_1$.

Then we have exactly one bit of single node information, and no higher order terms. 
Geometrically, $k$ lies in $\e_1$, so the only nonzero divergence in equation \eqref{series} is $d_1(k)$.
As one would expect, $d_2(k)$ and $d_3(k)$ vanish, as there is no synergy of order 2 and 3.

\begin{center}
 \includegraphics[scale=0.7,keepaspectratio=true]{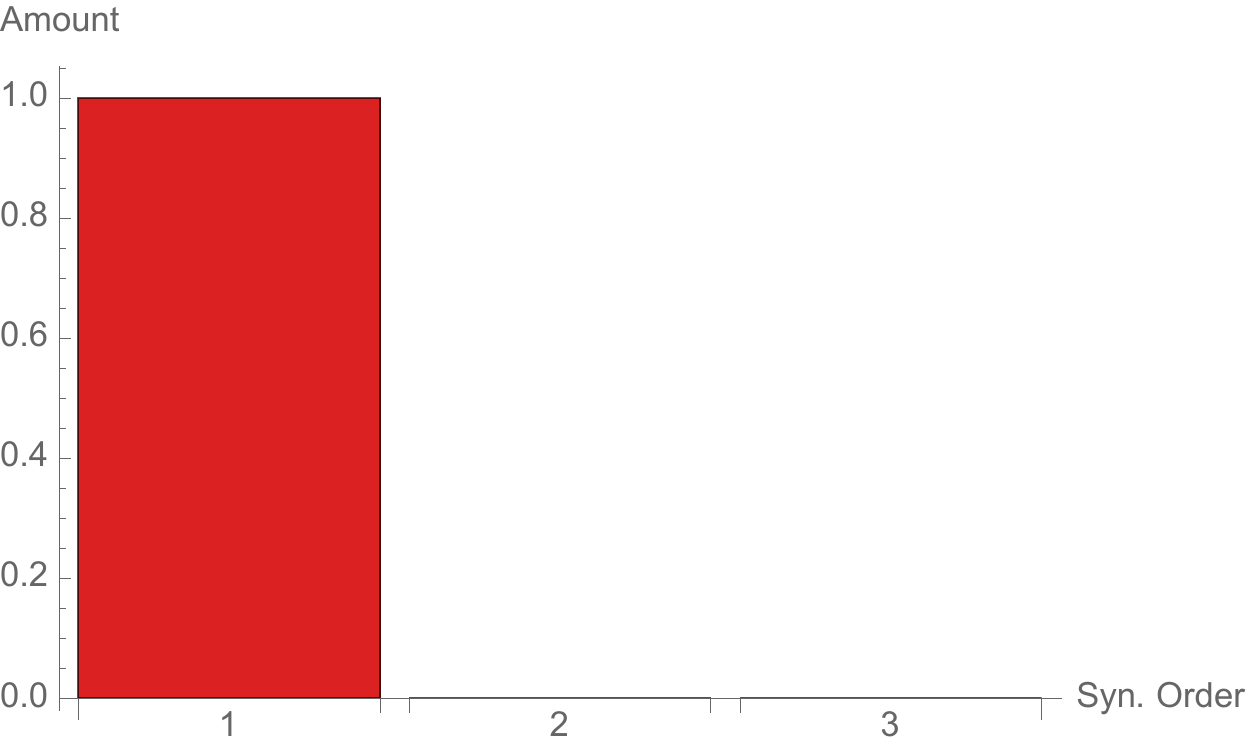}
\end{center}

\paragraph{Split Channel.}

The second easiest example is a more general channel which obeys equation \eqref{eone}.
In particular, consider 3 binary input nodes $X_1,X_2,X_3$ with constant input probability (so, the $x_i$ are independent), 
and output $Y=X_1\times X_2\times X_3$. As channel we simply take the identity map 
$(x_1,x_2,x_3)\mapsto(x_1,x_2,x_3)\in Y$. In this particular case:
\begin{equation}
 I(X:Y)=\sum_{i}I(X_i;Y)\;.
\end{equation}
We have 3 bits of mutual information,
which are all single node (but from different nodes). Since:
\begin{equation}
 k(x_1,x_2,x_3;y)=\phi_1(x_1,y_1)\,\phi_2(x_2,y_2)\,\phi_3(x_3,y_3)\;,
\end{equation}
which is a special case of \eqref{eone}, $k\in\e_1$, and so $d_2(k)$ and $d_3(k)$ in equation \eqref{series} 
are again zero.

\begin{center}
 \includegraphics[scale=0.7,keepaspectratio=true]{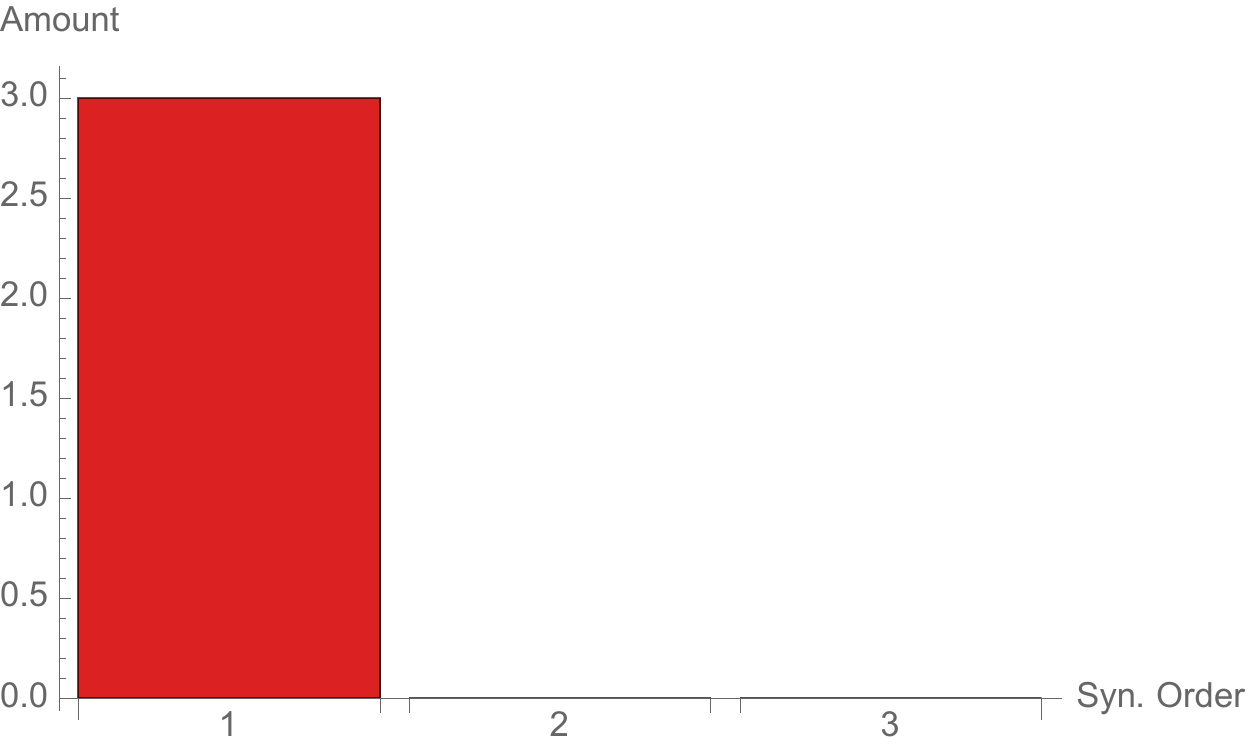}
\end{center}

\paragraph{Correlated Inputs.}

Consider 3 perfectly correlated binary nodes, each one with uniform marginal probability. 
As output take a perfect copy of one (hence, all) of the inputs. We have again one bit of mutual information,
which could come from any of the nodes, but no synergy, as no two nodes are interacting in the channel. 
The input distribution has correlation, but this has no effect on the channel, since the channel is simply copying 
the value of $X_1$ (or $X_2$ or $X_3$, equivalently). Therefore again $k\in\e_1$. Of the terms in equation \eqref{series},
again the only non-zero is $d_1(k)$.

\begin{center}
 \includegraphics[scale=0.7,keepaspectratio=true]{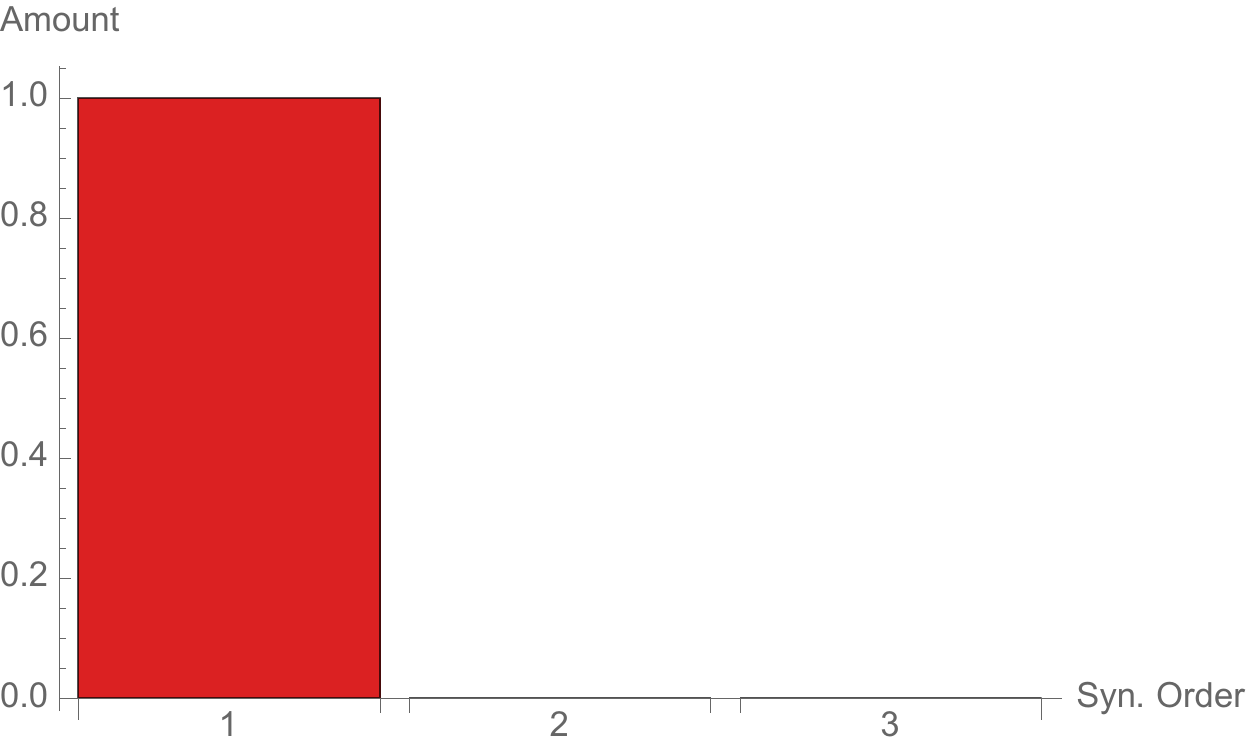}
\end{center}

This example in the literature is used to motivate the notion of redundancy.
A ``redundant channel'' is in our decomposition exactly equivalent to a single node channel, since it contains
exactly the same amount of information.

\paragraph{Parity (XOR).}

The standard example of synergy is given by the XOR function, and more
generally by the parity function between two or more nodes. 

For example, consider 3 binary input nodes $X_1,X_2,X_3$ with constant input probability, 
and one binary output node $Y$ which is given by $X_1\xor X_2$. We have 1 bit of mutual information,
which is purely arising from a pairwise synergy (of $X_1$ and $X_2$), so this time $k\in\e_2$.
The function XOR is \emph{pure} synergy, so $d_2(k)$ is the only non-zero term in \eqref{series}.

\begin{center}
 \includegraphics[scale=0.7,keepaspectratio=true]{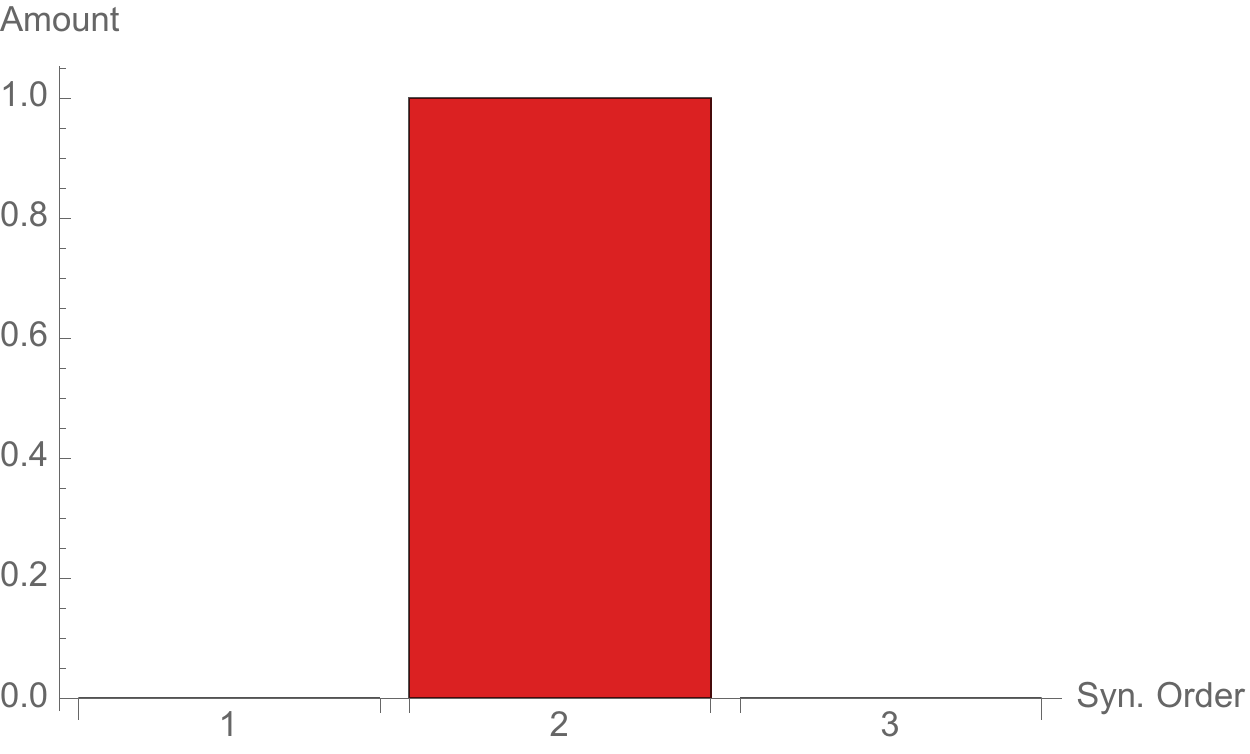}
\end{center}

If instead $Y$ is given by the threewise parity function, or $X_1\xor X_2\xor X_3$, we have again 1 bit of mutual information,
which now is purely arising from a threewise synergy, so here $k\in\e_3$, and the only non-zero term in \eqref{series} is $d_3(k)$.

\begin{center}
 \includegraphics[scale=0.7,keepaspectratio=true]{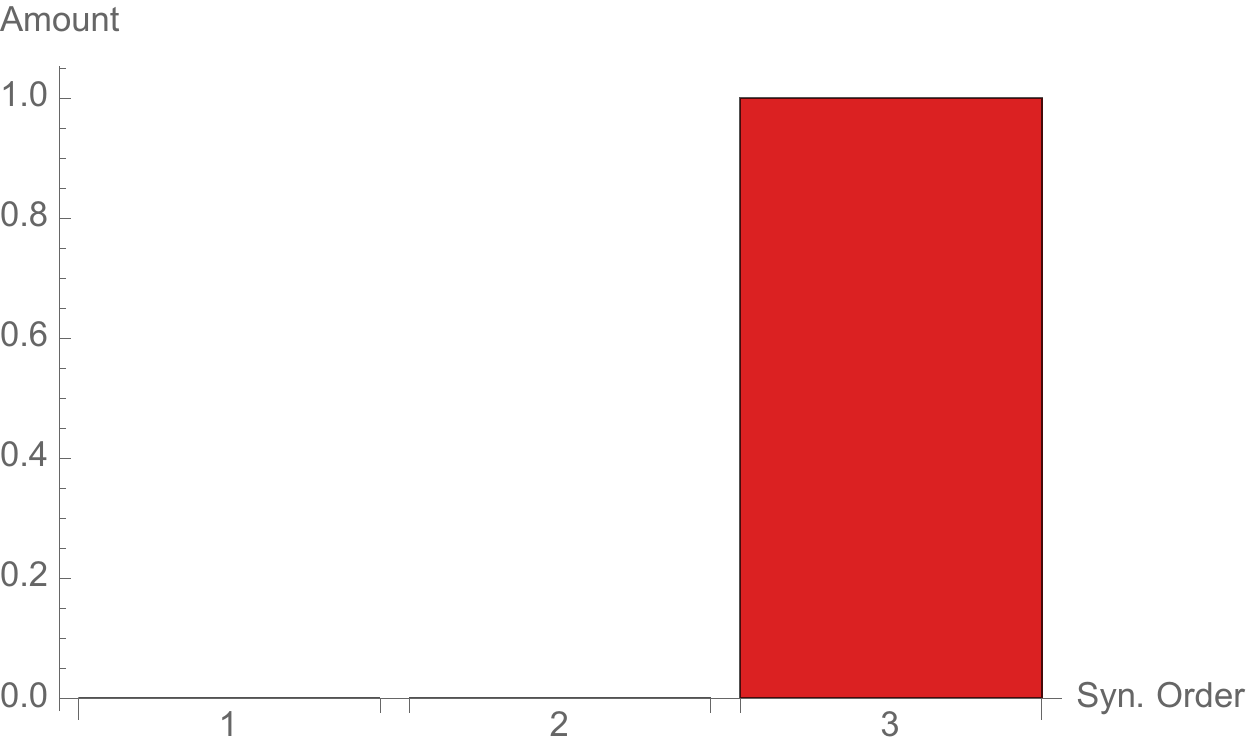}
\end{center}

In these examples there are no terms of lower order synergy, but the generic elements of $\e_2$ and $\e_3$
usually do contain a nonzero lower part. Consider for instance the next example.

\paragraph{AND and OR.}

The other two standard logic gates, AND and OR, share the same decomposition. 
Consider two binary nodes $X_1,X_2$ with uniform probability, 
and let $Y$ be $X_1\vee X_2$ (or $X_1\wedge X_2$). 
There is again one bit of mutual information, which comes mostly from single nodes, but also from
sy\-nergy. 

\begin{center}
 \includegraphics[scale=0.7,keepaspectratio=true]{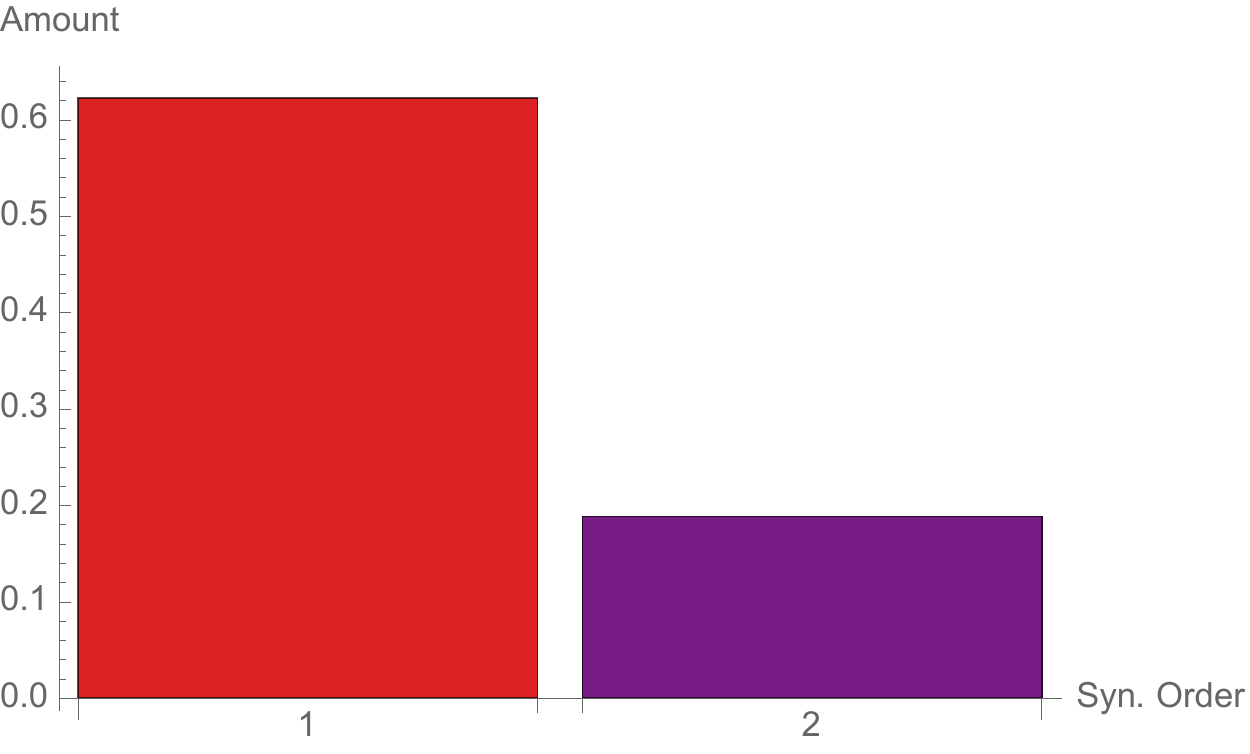}
\end{center}

Geometrically, this means that AND and OR are channels in $\e_2$ which lie close to the submanifold $\e_1$.

\paragraph{XorLoses.}

Here we present a slightly more complicated example, coming from \cite{griffith}. We have three binary nodes 
$X_1,X_2,X_3$, where $X_1,X_2$ have uniform probabilities, and an output node $Y=X_1\xor X_2$, just like 
in the ``XOR'' example. Now we take $X_3$ to be perfectly correlated with $Y=X_1\xor X_2$, 
so that $Y$ could get the information either from $X_3$ or from the synergy between
$X_1$ and $X_2$. We have one bit of mutual information, which can be seen as entirely coming from $X_3$,
and so the synergy between $X_1$ and $X_2$ is not adding anything.

\begin{center}
 \includegraphics[scale=0.7,keepaspectratio=true]{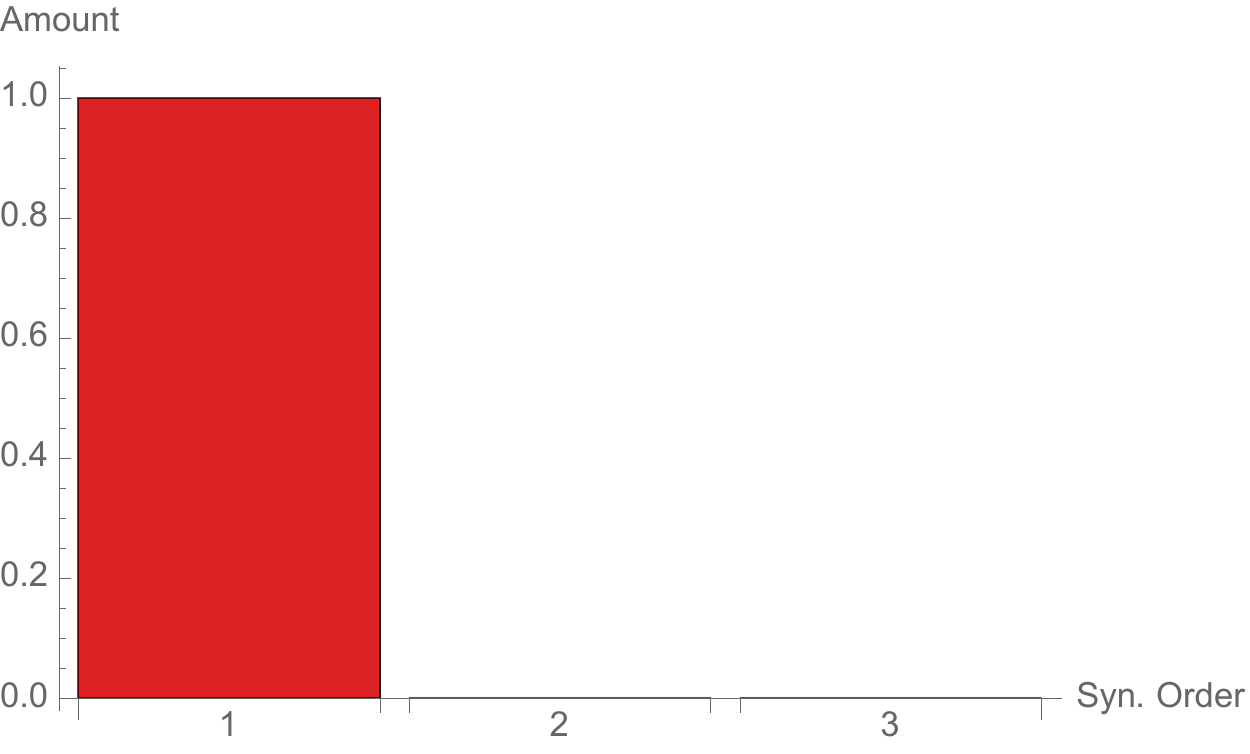}
\end{center}

\paragraph{XorDuplicate.}

Again from \cite{griffith}. We have 3 binary nodes 
$X_1,X_2,X_3$, where $X_1,X_2$ have uniform probabilities, while $X_3=X_1$. The output is
$X_1\xor X_2=X_3\xor X_2$, so it could get the information either from the synergy between
$X_1$ and $X_2$, or $X_2$ and $X_3$. There is one bit of mutual information, which is coming from a pairwise
interaction. Again, it does not matter between whom. 

\begin{center}
 \includegraphics[scale=0.7,keepaspectratio=true]{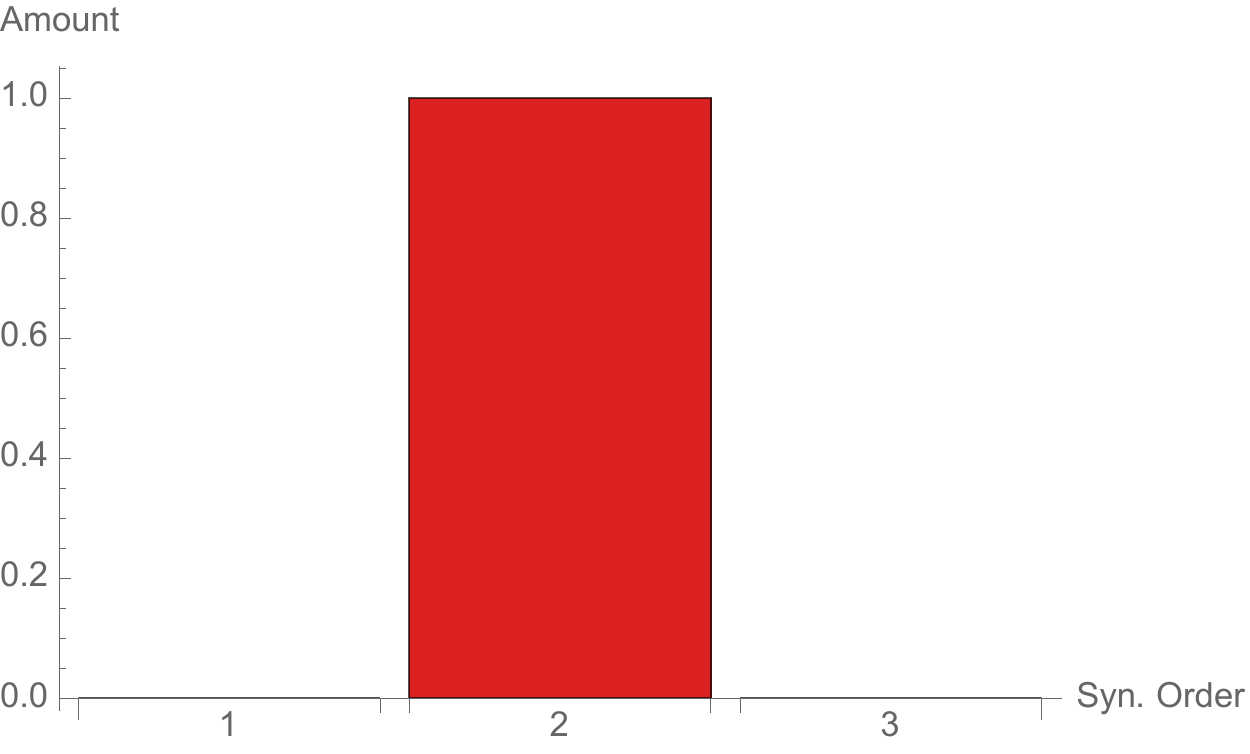}
\end{center}

It should be clear from the examples here that decomposing only \emph{by order}, and not by the specific subsets, is crucial. For example, 
in the ``input correlation'' example, there is no natural way to decide from \emph{which single node} the information comes, even if it is clear 
that the interaction is of order 1.

\section{General Case}\label{decomp}

Here we try to give a general formulation, for $N$ inputs, of the quantities defined in Section \ref{idea}. 
As in the introduciton we call the set of input nodes $V$ of cardinality $N$, and we consider a subset of
the nodes $I$. We denote the joint random variable $(X_i,i\in I)$ by $X_I$, and we denote the complement of $I$ 
in $V$ by $I^c$.
The case $N=3$ in Section \ref{idea} should motivate the following definition.

\begin{deph}
 Let $I\subseteq V$. We call $F_I$ the space of functions who only depend on $X_I$ and $Y$:
 \begin{equation}
  F_I:= \big\{ f\in F(X,Y)\;\big|\;f(x_I,x_{I^c};y)=f(x_I,x'_{I^c};y)\;\forall x_{I^c}, x'_{I^c} \big\}\;.
 \end{equation}
 Let $0\le i \le N$. We call $F_i$ the space of channels which can be written as a product of functions 
 of $F_I$ with the order of $I$ at most $k$:
 \begin{equation}
  \e_i:= cl\bigg\{ k\in K(X;Y)\;\bigg|\;\exists \phi_I\in F_I,\phi_0\in F(X) \,\big|\, k=\phi_0\prod_{I}\phi_I;\;|I|\leq i \bigg\}\;,
 \end{equation}
 where $cl$ denotes the closure in $K(X;Y)$. Intuitively, this means that $\e_i$ does not only contain terms in the 
 form given in the curly brackets, but also limits of such terms. 
 Stated differently, the closure of a set includes not only the set itself, but also its boundary. This
 is important, because when we project to a family, the projection may lie on the boundary. In order for the 
 result to exist, the boundary must then be included. 
\end{deph}

This way:
\begin{itemize}
 \item $\e_0$ is the space of constant channels;
 \item $\e_N$ is the whole $K(X;Y)$;
 \item $\e_i\subseteq \e_j$ if and only if $i\leq j$;
 \item For $N\leq3$ we recover exactly the quantities of Section \ref{idea}.
\end{itemize}

The family $\e_i$ is also the closure of the family in the form:
\begin{equation}
 \Bigg\{ \dfrac{1}{Z(X)} \exp\left( \sum_{I}f_I(X;Y) \right)\bigg|\;f_I\in F_I;\;|I|\leq i \Bigg\}\;,
\end{equation}
where:
\begin{equation}
 Z(x):=\sum_y\exp\left( \sum_{I}f_I(x;y) \right)\;.
\end{equation}
Such families are known in the literature as \emph{exponential families} (see for example \cite{amari}).
In particular, it is compact (for finite $N$), so that the infimum of any function on $\e_i$ is always a minimum.
This means that for a channel $k$ and an input distribution $p$:
\begin{equation}
 D_p(k||\e_i) := \inf_{m\in\e_i}D_p(k||m) = \min_{m\in\e_i}D_p(k||m)
\end{equation}
always exists. If it is unique, for example if $p$ is strictly positive, we define the unique KL-projection as:
\begin{equation}
 \pi_{\e_i}k:=\argmin_{m\in\e_i}D_p(k||m)\;.
\end{equation}
$\pi_{\e_i}k$ has the property that it defines the same output probability on $Y$.

\begin{deph}
 Let $k\in K(X;Y)$, let $1\leq i \leq N$. Then the $i$-wise synergy of $k$ is (if the KL-projections are unique):
 \begin{equation}
  d_i(k):= D_p(\pi_{\e_i}k||\pi_{\e_{i-1}}k)\;.
 \end{equation} 
 For more clarity, we call the $1$-wise synergy ``single node information'' or ``single-node dependence''. 
\end{deph}

For $k\in K(X;Y)=\e_N$, we can look at its divergence from $\e_0$.
If we denote $\pi_{\e_0}k$ by $k_0$:
\begin{equation}\label{min}
 D_p(k||\e_0) = D_p(k||k_0) = \sum_{x,y} p(x)\,k(x;y)\,\log \dfrac{k(x;y)}{k_0(y)}\;.
\end{equation}
If $k$ is not strictly positive, we take the convention $0\log(0/0)=0$, 
and we discard the zero terms from the sum. Since ${k_0}_*p=k_*p$ but $k_0$ is constant in $x$, it can \emph{not} happen
that for some $(x;y)$, $k_0(x;y)=0$ but $k(x;y)\ne0$.  
(The very same is true for all KL-projections $\pi_{\e_i}k$, since $D_p(\pi_{\e_i}k||0)\leq D_p(k||k_0)$.)
For all other terms, \eqref{min} becomes:
\begin{align}
 D_p(k||\e_0) &= \sum_{x,y}p(x)\, k(x;y) \log k(x;y) - \sum_{x,y} p(x)\, k(x;y) \log k_0(y) \\
  &=-H_{pk}(Y|X) - \sum_y k_*p(y) \log k_0(y) \\
  &= -H_{pk}(Y|X) - \sum_y {k_0}_*p(y) \log k_0(y) \\
  &= -H_{pk}(Y|X) + H_{{k_0}_*p}(Y)  = -H_{pk}(Y|X) + H_{{k}_*p}(Y) \\
  &= I_{pk}(X:Y)\;.
\end{align}
On the other hand, the Pythagorean relation \eqref{pyt} implies:
\begin{equation}
 D_p(k||k_0) = D_p(k||\pi_{\e_{N-1}}k) + D_p(\pi_{\e_{N-1}}k||k_0)\;,
\end{equation}
and iterating:
\begin{equation}
 D_p(k||k_0) = D_p(k||\pi_{\e_{N-1}}k) + D_p(\pi_{\e_{N-1}}k||\pi_{\e_{N-2}}k)+\dots+D_p(\pi_{\e_{1}}k||k_0)\;.
\end{equation}
In the end, we get:
\begin{equation}\label{deko}
 I(X:Y)= \sum_{i=1}^N D_p(\pi_{\e_{i}}k||\pi_{\e_{i-1}}k) = \sum_{i=1}^N d_i(k)\;.
\end{equation}

This decomposition is always non-negative, and it depends on the input distribution. 
The terms in \eqref{deko} can be in general difficult to compute exactly. Nevertheless, they can be approximated
with iterative procedures.

\section{Comparison with the Measure of \cite{griffith} and \cite{entropy}}\label{jfr}

The measure of synergy, or respectively complementary information, defined in \cite{griffith} and \cite{entropy},
is:
\begin{equation}\label{ci}
 CI_p(Y:X_1,X_2):= I_p(Y:X_1,X_2)-\min_{p^*\in \wedge} I_{p^*}(Y:X_1,X_2)\;,
\end{equation}
where $\wedge$ is the space of prescribed marginals:
\begin{equation}
 \wedge = \big\{ q\in P(X_1,X_2,Y)\;\big|\; q(X_1,Y)=p(X_1,Y), q(X_2,Y)=p(Y_2,Y) \big\}\;.
\end{equation}
Our measure of synergy can be written, for two inputs, in a similar form:
\begin{align}
 d_2(k) &= D_p(k||\pi_{\e_1}k) = =I_p(Y:X_1,X_2)-\min_{p^*\in \vartriangle} I_{p^*}(Y:X_1,X_2)\;,
\end{align}
where $\vartriangle$, in addition to the constraints of $\wedge$, prescribes also the input:
\begin{align}
 \vartriangle\, = \big\{ &q\in P(X_1,X_2,Y)\;\big|\\ 
  \notag &q(X_1,Y)=p(X_1,Y), q(X_2,Y)=p(Y_2,Y), q(X_1,X_2)=p(X_1,X_2) \big\}\;.
\end{align}
Clearly $\vartriangle\,\subseteq\wedge$, so:
\begin{equation}
  \min_{p^*\in \vartriangle} I_{p^*}(Y:X_1,X_2) \ge \min_{p^*\in \wedge} I_{p^*}(Y:X_1,X_2) \;,
\end{equation}
which implies that:
\begin{equation}
 d_2(k) \le CI_p(Y:X_1,X_2)\;.
\end{equation}

We argue that not prescribing the input leads to overestimating synergy, because 
the subtraction in \eqref{ci} includes a possible difference in the correlation of the input distributions. 

For example, consider $X_1,X_2,Y$ binary and correlated, but not \emph{perfectly} correlated. 
(For perfectly correlated nodes, as in Section \ref{examples}, $\vartriangle\,=\wedge$, so
there is no difference between the two measures.)
In detail, consider the channel:
\begin{equation}
 k_\beta(x_1,x_2;y):=\dfrac{\exp\left( \beta\, y\,(x_1+x_2) \right)}{\sum_{y'}\exp\left( \beta\, y'(x_1+x_2) \right)}\;,
\end{equation}
and the input distribution:
\begin{equation}
 p_\alpha(x_1,x_2):= \dfrac{\exp\left( \alpha\,x_1 x_2 \right)}{\sum_{x'_1,x'_2}\exp\left( \alpha\, x'_1 x'_2 \right)}\;.
\end{equation}
For $\alpha,\beta\to\infty$, the correlation becomes perfect, and the two measures of synergy are both zero. 
For $0<\alpha,\beta<\infty$, our measure $d_2(k_\beta)$ is zero, as clearly $k_\beta\in\e_1$.
$CI$ is more difficult to compute, but we can give a (non-zero) lower bound in the following way. 
First we fix two values $\beta=\beta_0,\alpha=\alpha_0$. We consider the joint distribution $p_{\alpha_0}k_{\beta_0}$, 
and look at the marginals:
\begin{equation}
 p_{\alpha_0}k_{\beta_0} (X_1,Y)\;,\quad  p_{\alpha_0}k_{\beta_0} (X_2,Y)\;.
\end{equation}
We define the family $\wedge$ as the set of joint probabilities which have exactly these marginals.
If we increase $\beta$, we can always find an $\alpha$ such that the marginals do not change:
\begin{equation}
 p_{\alpha}k_{\beta} (X_1,Y)= p_{\alpha_0}k_{\beta_0} (X_1,Y)\;,\quad  p_{\alpha}k_{\beta} (X_2,Y)=p_{\alpha_0}k_{\beta_0} (X_2,Y)\;,
\end{equation}
i.e. such that $p_\alpha k_{\beta}\in\wedge$.
Now we can look at the mutual information of $p_{\alpha}k_{\beta}$ and of $p_{\alpha_0}k_{\beta_0}$. If they
differ, and (for example) the former is larger, then:
\begin{align}
 I_{p_{\alpha}k_{\beta}}(Y:X_1,X_2) &- I_{p_{\alpha_0}k_{\beta_0}}(Y:X_1,X_2) \notag \\
  &\le I_{p_{\alpha}k_{\beta}}(Y:X_1,X_2)- \min_{p^*\in\wedge}I_{p^*}(Y:X_1,X_2) = CI_{p_{\alpha}k_{\beta}}
\end{align}
is a well-defined lower bound for $CI_{p_{\alpha}k_{\beta}}$.
With a numerical simulation we can show graphically that the mutual information is indeed not constant within 
the families $\wedge$.

\begin{figure}[H]\label{impl}
 \centering
 \includegraphics[scale=0.7,keepaspectratio=true]{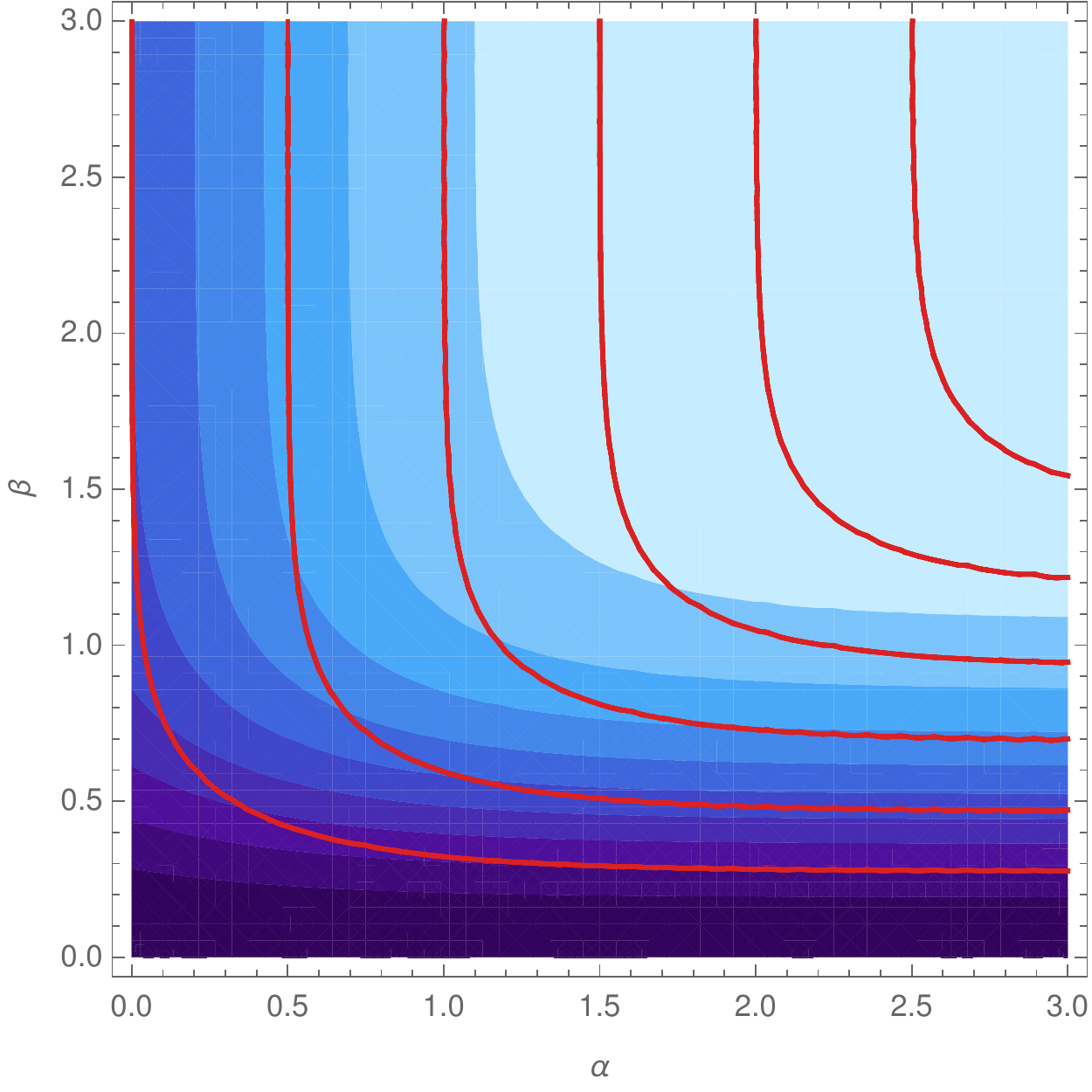}
 \caption{Mutual information and fixed marginals. The shades of blue represent the amount of $I_p(Y:X_1,X_2)$ 
 as a function of $\alpha,\beta$ (brighter is higher). Each red line represents a family $\wedge$ of fixed marginals. 
 While the lines of fixed mutual information and the families of fixed marginals look qualitatively similar,
 they do not coincide exactly, which means that $I_p$ varies within the $\wedge$.}
\end{figure}

From the picture we can see that the red lines (families $\wedge$ for different initial values) approximate well the lines of constant 
mutual information, at least qualitatively, but they are not exactly equal. This means that for most points $p$ of
$\wedge$, the quantity:
\begin{equation}
 CI_p(Y:X_1,X_2):= I_p(Y:X_1,X_2)-\min_{p^*\in \wedge} I_{p^*}(Y:X_1,X_2)
\end{equation}
will be non-zero.
More explicitly, we can plot the increase in mutual information as $p$ varies in $\wedge$, for example as a 
function of $\beta$. This is always larger or equal than the the difference between the mutual information 
and its minimum in $\wedge$ (i.e. $CI$). We can see that it is positive, which implies that $CI_p$ is also positive.

\begin{figure}[H]\label{expl}
 \centering
 \includegraphics[scale=0.7,keepaspectratio=true]{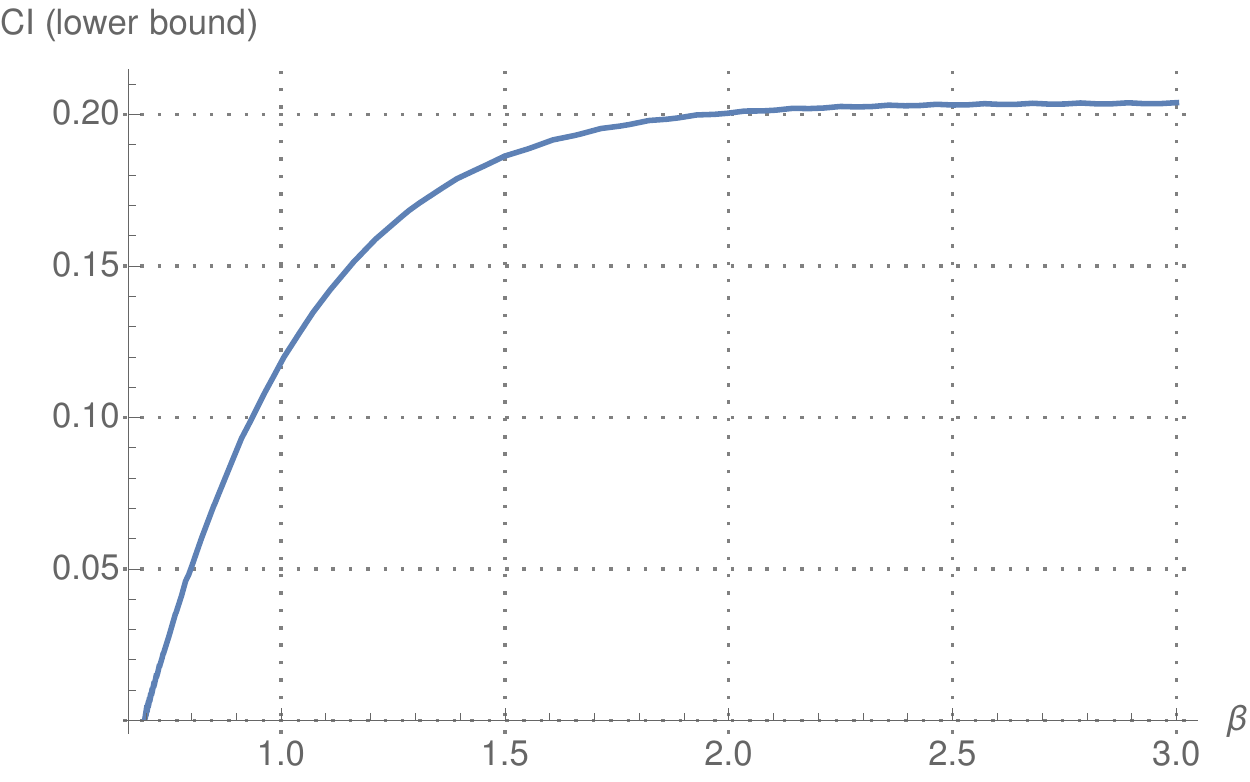}
 \caption{Lower bound for $CI$ versus $\beta$. 
 For each $\beta\in[0.7,3]$ we can find an $\alpha$ such that the joint $p_\alpha k_{\beta}$ lies in $\wedge$. 
 The increase in mutual information as $\beta$ varies is a lower bound for $CI$, which is then in general non-zero.}
\end{figure}

We can see in Figure 3 that, especially for very large or very small values of $\alpha$ and $\beta$ 
(i.e. very strong or very weak correlation), $CI$ captures the behaviour of mutual information quite well. 
These limits are precisely deterministic and constant kernels, for which most approaches in quantifying synergy coincide.
This is the reason why the examples studied in \cite{griffith} give quite a satisfying result for $CI$ (in
their notation, $S_{VK}$).
For the less studied (and computationally more complex) 
intermediate values, like $1<\alpha,\beta<2$, the approximation is instead far from accurate, and in
that interval (see Figure 4) there is a sharp increase in $I$, which leads to overestimating synergy.

\section{Conclusion}

Using information geometry, we have defined a non-negative decomposition of the mutual information between 
inputs and outputs of a channel. 

The decomposition divides the mutual information into contributions of the different orders of synergy in a unique way. 
It does \emph{not}, however, divide the mutual information into contributions of the diffe\-rent subsets of input nodes as 
Williams and Beer's PID \cite{beer} would require. 

For two inputs, our measure of synergy is closely related to the well-received quantification of synergy in \cite{griffith} and \cite{entropy}. Our measure though
works in the desired way for an arbitrary (finite) number of inputs. Differently from \cite{griffith} and \cite{entropy}, anyway, we do not define a measure
for redundant or ``shared'' information, nor unique information of the single inputs or subsets. 

The decomposition depends on the choice of an input distribution. 
In case of input correlation, redundant information is counted automatically only once. This way there is no need to quantify redundancy separately.

In general there is no way to compute our quantities in closed form, but they can be approximated by an iterative scaling algorithm (see for example \cite{csiszar}). 
The results are consistent with the intuitive properties of synergy, outlined in \cite{beer} and \cite{griffith}.

\end{document}